# Assessment of Generative Adversarial Networks Model for Synthetic Optical Coherence Tomography Images of Retinal Disorders

Ce Zheng[1], Xiaolin Xie[4], Kang Zhou[2,3], Bang Chen[2], Jili Chen[5], Haiyun Ye[1], Wen Li[1], Tong Qiao[1], Shenghua Gao[3], Jianlong Yang[2,*], Jiang Liu[2,6]

1. Department of Ophthalmology, Shanghai Children's Hospital, Shanghai Jiao Tong University, Shanghai, China

2. Ningbo Institute of Industrial Technology, Chinese Academy of Sciences, China

3. Shanghai Tech University, China

4. Joint Shantou International Eye Center of Shantou University and the Chinese University of Hong Kong, Shantou University Medical College, Shantou, Guangdong, China.

5. Department of Ophthalmology, Shibei Hospital, Shanghai, China

6. Southern University of Science and Technology, China

Emails: Ce Zheng (zhengce@me.com); Xiaolin Xie (xiexl@jsiec.org); Tong Qiao (qiaojoel@163.com); Kang Zhou (zhoukang@shanghaitech.edu.cn); Bang Chen (chenbang@nimte.ac.cn); Jili Chen (corneachen@163.com); Haiyun Ye (yehy@shchildren.com.cn); Wen Li (liwen@shchildren.com.cn); Shenghua Gao (gaoshh@shanghaitech.edu.cn); Jiang Liu (liuj@sustech.edu.cn).

* Corresponding Author:   Jianlong Yang (yangjianlong@nimte.ac.cn )

Ningbo Institute of Industrial Technology, Chinese Academy of Sciences, China

Telephone: +86-158-0194-9951

**Abstract**

**Purpose:** To assess whether a generative adversarial network (GAN) could synthesize realistic optical coherence tomography (OCT) images that satisfactorily serve as the educational images for retinal specialists and the training datasets for the classification of various retinal disorders using deep learning (DL).

**Methods:** The GANs architecture was adopted to synthesis high-resolution OCT images training on a publicly available OCT dataset including urgent referrals (choroidal neovascularization and diabetic macular edema) and non-urgent referrals (normal and drusen). 400 real and synthetic OCT images were evaluated by 2 retinal specialists to assess image quality. We further trained 2 DL models on either real or synthetic datasets and compared the performance of urgent vs non-urgent referrals diagnosis tested on a local (1000 images from the public dataset) and clinical validation dataset (278 images from Shanghai Shibei Hospital).

**Results:** The image quality of real vs synthetic OCT images was similar as assessed by 2 retinal specialists. The accuracy of discrimination as real vs synthetic OCT images was 59.50% for retinal specialist 1 and 53.67% for retinal specialist 2. For the local dataset, the DL model trained on real (DL_Model_R) and synthetic OCT images (DL_Model_S) had an area under the curve (AUC) of 0.99, and 0.98 respectively. For the clinical dataset, the AUC was 0.94 for DL_Model_R, 0.90 for DL_Model_S.

**Conclusions:** The GAN-synthetic OCT images can be used by clinicians for educational purposes and developing DL algorithms.

**Translational Relevance:** The medical image synthesis based on GANs is promising in human and machine to fulfill clinical tasks.

**INTRODUCTION:**

Optical coherence tomography (OCT), which uses laser to capture high-resolution retina images in vivo [1], is now a standard of care for guiding the diagnosis and treatment of some of the leading causes of blindness worldwide, including age-related macular degeneration (AMD) and diabetic macular edema (DME) [2-3]. Given the increasing prevalence of these diseases, deep learning (DL) algorithms has be utilized as alternative screening tools to rectify the manpower and expertise shortage [4-6]. Kermany et al., recently demonstrated highly accurate DL algorithms for OCT imaging classification and performance is comparable to that of human experts [7].

Despite these promising results, DL algorithm require large, diverse, and well-balanced image training data sets with labels defining structures [8]. For example, Kermany et al., trained a DL algorithm using a training dataset with totally 108,312 images by sharing data from different centers [7]. Such approach has several limitations. First, when data are to be shared between different centers regulations and state privacy rules need to be considered. As defined by National Institute of Standard and Technology (NIST, USA), biometric data including retina images is kind of personally identifiable information (PII) and could possibly be protected from inappropriate access regardless of the original individual study participant consent or local institutional review board permission [9]. Moreover, larger datasets do not necessarily enhance the performance of a DL algorithm. For example, adding large amounts of unbalanced data, like images from healthy subjects, will most likely not improve performance.

In order to address these disadvantages, several authors suggested using generative adversarial networks (GANs) to synthesize new images from a training data set of real images [10]. Using Age-Related Eye Disease Study (AREDS) dataset of 133, 821 fundus images, Burlina et al., generated similar number of synthetic images to train DL model [11]. They reported DL models trained with only synthetic images showed performance nearing that resulting from training on real images. Therefore, the aim of this study was to assess whether a GAN neural network could synthesize realistic OCT images that satisfactorily serve as training datasets for DL algorithms and education images for retinal specialists.

**METHODS:**

*Datasets*

Approval was obtained from the institutional review board (IRB) of Shanghai Children's Hospital (identifier, 2018RY029-E01) and Shanghai Shibei Hospital (identifier, YL_201805258-05) to conduct the study in accordance with the tenets of the Declaration of Helsinki. The informed consent was not required due to the anonymized usage of images and the retrospective study design.

In current study, we used OCT images from Kermany et al's datasets (abbreviated as Cell's dataset). The detail of this study and Cell's dataset had been described previously [7]. Briefly, a DL model was trained using a dataset with totally 108,312 images after a tiered grading system. The whole training datasets included 37,206 with choroidal neovascularization, 11,349 with diabetic macular edema, 8,617 with drusen, and 51,140 normal respectively. The study further categorized images with choroidal neovascularization (CNV) and diabetic macular edema (DME) as ''urgent referrals.'', drusen and normal as "non-urgent referrals". In original study, 1,000 images (250 from each category) were used as a local validation set. To test the generalization of synthetic OCT images and DL models, a second clinical testing set was collected from department of ophthalmology, Shanghai Shibei Hospital (SSH), from July 2018 to Feb 2019. After searching local electronic medical record databases, totally 278 OCT images were downloaded using a standard image format based on the manufacturer's instructions (Heidelberg Spectralis, Heidelberg Engineering, Germany). The clinical dataset comprised 130 OCT images with urgent referrals (CNV and DME) and 148 OCT images with non-urgent referrals (drusen and normal) respectively.

*Development and Evaluation of Generative Adversarial Networks (GAN) Synthetic OCT Images*

We adopted progressively grown Generative Adversarial Networks (PGGANs) to synthesis high-resolution OCT images [12]. GANs consist a discriminator network (D) and a generative network (G), which is trained by adversarial learning strategy. With the adversarial learning between the G and the D, the generative network is promoted to generate fake with much more similarity to the real images. PGGANs is an extension to the GAN training process and achieve high-resolution synthetization (in this study, 256 × 256 pixels) by alternating training and adding new network G and D. To generate OCT images with correct anatomical structures, a sketch guidance modules G that contains the edge and detail information was added to growing networks G [13]. With each additional layer, the resolution is increased (eg, from 4 × 4 to 8 × 8) allowing the generation of higher-resolution images. All of the GAN development and experiments were conducted using PyTorch (version 1.0, Facebook, USA). The GANs were trained on an Ubuntu 16.04 operation system with Intel Core i7-2700K 4.6 GHz CPU, 128 GB RAM, and NVIDIA GTX 1080Ti 12 GB GPU. In this study, GANs took about 40 hours to generate 100,456 synthetic OCT images, including 48,751 images as urgent referrals and 51,705 images as non-urgent referrals respectively.

To assess whether synthetic OCT images appear realistic to human experts and can be used for clinical evaluation, 400 OCT images (equal numbers of real and synthetic images) were evaluated by 2 human experts to assess image quality and gradability of real and synthetic OCT images. Human experts included 2 retinal specialists who had more than 10 (retinal specialist 1; HYY.) and 20 (retinal specialist 2; WL.) years of clinical experience respectively.

Firstly, they were asked to determine whether the image quality is sufficient for clinical grading. Once finished grading, they were informed that image set composed of a mixture of real and synthetic images and asked to determine whether image is real or synthetic.

*Evaluation of GAN Synthetic OCT Images Used for DL Classification*

To assess whether synthetic OCT images can be used as training dataset for DL model, we evaluated diagnostic performance of urgent vs. non-urgent referable classification by comparing 2 DL models trained on real (DL_Model_R) and synthetic (DL_Model_S) OCT images respectively. Transfer learning with fine-tune technique was adopted to build the DL models by using a modified Inception V3 architecture with weights pre-trained on ImageNet [14,15]. After removing the final classification layer from the network, we further retained DL models with real and synthetic OCT images independently. DL models were implemented in Tensorflow framework (Google, version 1.10.0) with Keras API (version 2.2.4). All images were resized to 299*299 pixels as required by Keras' API. Data augmentation was performed to increase the amount and type of variation within the training dataset, including included horizontal flipping, rotation of $10°$, and sharpening and adjustments to saturation, brightness, contrast, and color balance. Training was then performed by a stochastic gradient descent in minibatch size of 32 images with an Adam optimizer learning rate of 0.001. Training was run for 200 epochs, as the absence of further improvement in both accuracy and cross-entropy loss (Figure 2).

To test the generalization of DL models, we assessed of the DL models in 2 different test datasets. The first local validation dataset composed of same testing dataset from Cell's dataset and the second clinical validation dataset collected from SSH. The DL models' classification performance in both local and clinical validation dataset was then compared for DL-R and DL-S.

*Statistics*

The performance of our algorithm was evaluated in terms of area under the receiver operating characteristic curve (AUC), precision, recall and f1 score with 2-sided 95% CIs. On. The formulas for calculating the Acc, precision, recall and f1 score were defined as:

$$\text{Acc} = \frac{TP+TN}{TP+TN+FN+FP}, \tag{1}$$

$$\text{Precision} = \frac{TP}{TP+FP}, \tag{2}$$

$$\text{Recall} = \frac{TP}{TP+FN}, \tag{3}$$

$$\text{f1} = \frac{2\times(Precision\times Recall)}{Precision+Recall}, \tag{4}$$

where TP, TN, FP, and FN are the true positives, true negatives, false positives, and false negatives, respectively.

All statistical analyses were performed using the Python and Scikit_learn modules (Anaconda Python, Continuum Analytics).

**RESULTS:**

*Image Quality and Discrimination between Real and Synthesizing OCT Images by Human Experts*

Figure 2 demonstrates some examples of the synthetic OCT images. Overall, our approach is capable of generating OCT images that are realistic. The results of image quality graded by 2 retinal specialists are shown in Table 1. Retinal specialist 1 rated 1.5% of real images to be poor quality vs 2% of synthetic images. Retinal specialist 2 rated 2% of real images to be poor quality vs 2.5% of synthetic images. Our results revealed that both real and synthetic OCT images have approximately same quality for human experts.

When comparing discrimination between real vs synthetic OCT images, our results showed that 2 human experts had limited ability to discern real from synthetic images, with an accuracy of 59.50% (95% CI: 53.5% - 65.3%) for retinal specialist 1 and 53.67% (95% CI: 47.8% - 59.5%) for retinal specialist 2.

*Performance of DL Classification on Real and Synthesizing OCT Images*

The diagnostic performance of 2 DL models, testing in local Cell validation and clinical validation datasets, are shown as confusion matrices in Fig. 3 and Fig. 4. For local Cell validation dataset, the AUC was 0.99 for DL_Model_R (with 95% CI, 0.99 to 1.00), 0.98 for DL_Model_S (with 95% CI, 0.99 to 1.00). For clinical validation dataset, the AUC was 0.94 for DL_Model_R (with 95% CI, 0.91 to 0.97), 0.90 for DL_Model_S (with 95% CI, 0.91 to 0.97) as listed in Table 2. Figure 5 is their ROC curves. Overall, moderate performance decrease was achieved in both local and clinical validation dataset when using DL model trained only with synthetic OCT images, but the performance was still considered to be good.

**Discussion**

In this study, we developed and evaluated GANs adopted to generating high-resolution OCT images with urgent and non-urgent referrals. The results suggest that synthetic OCT images can be used by clinicians in place of real images for education or clinical training purpose. Moreover, moderate performance decrease using the synthetic training data set, suggesting that synthetic OCT images can also be served as augmentation of training datasets for use by DL models.

Deep learning (DL) have make dramatical progress for discriminative medical image analysis tasks and achieved performance exceeding traditional machine learning and close to that of human experts [16-18]. However, DL approaches require a large number of high-quality data. An obvious approach is data sharing from different centers. This is often impeded by IRB concerns, patients' consent, or proprietary data restrictions. Moreover, the implementation of data sharing requires hardware and software investments, expertise and is labor-intensive. Recently, GANs are proposed to generate synthetic images that matches the real images via an adversarial process. GANs have been successfully applied to many medical image synthesis tasks, including retinal fundus, melanoma lesion, CT and MRI images synthesizing [19-21]. Burline et al suggested that GANs–synthesized fundus images of AMD are realistic and could be used for both education and for machine training [18]. Using similar GANs model mentioned by previous study, our results also shown the ability of GANs to synthesize realistic OCT images. This result was encouraging in that the GANs can generated synthetic OCT image with high quality assessed by human experts.

The potential application of this technique is promising, as our study also showed that GANs OCT images can also be served as image augmentation for both clinical purpose and DL models training. Previously, most of DL research groups used data sharing from different centers to increase the number of input data for network training. For detecting glaucomatous optic neuropathy, Liu et al. reported DL model with AUC of 0.996 training on 6 different eye centers with more than 200 thousand fundus images [5]. Using datasets from two different countries, Gulshan et al, reported AUC of 0.974 for detection of diabetic retinopathy [4]. Recently, Kermany et al reported a DL model with AUC of 0.999 for screening common treatable blinding retinal diseases, like CNV or EMD, using OCT images. This study involved more than 100 thousand OCT images in 4 eye centers from 2 different country. It is interesting to notice that, using same dataset for generating synthetic OCT image, our model trained on all-synthetic OCT images achieved similar AUC of 0.98 with that of DL model trained in all-real OCT images. Our result suggested that this technique also hold promise for augmenting DL models while preserving anonymity and obviating constraints owing to institutional review board or other use restrictions.

It is important to ensure the generalizability of DL model by testing in independent datasets with clinical settings [22]. As recently work showing highly accurate DL algorithms for various medical image classification tasks in well-labeled dataset, it is interesting to validate the performance of DL models in real life. Nishanthan, et al, reported a moderate decreasing of AUC (0.901 vs. 0.980) when using clinical validation dataset for screening diabetic retinopathy. In our study, moderate performance decrease was achieved in both local and clinical validation dataset when using DL model trained only with synthetic OCT images, but the performance was still considered to be good.

One limitation of this study was images synthesized in our study were $256 \times 256$ pixels, which is lower than was that of OCT images used in the original Ker' study. The default input size for inception BE is 299 × 299 pixels. It is possible for GANs to generate higher resolutions (e.g., 1024 × 1024 or above), however it required a longer training time (at least one month for the hard ware used in current study), which was impractical for this study. Second, other retinal disorders like macular hole, epiretinal membrane or pigment epithelium detachment were not included in current study due to less common condition. Future work will involve generating such rare retinal disorders and might help improve diagnostic accuracy of DL models

In summary, the GAN-synthetic OCT images can be used by clinicians in place of real OCT images for clinical purpose and developing DL algorithms.

**Acknowledgments**

**Funding:** This study was supported by the National Natural Science Foundation of China (81371010), Clinical Research Funds of Shantou University Medical College (2014), and Ningbo 3315 Innovation team grant, Cixi Institute of Biomedical Engineering, Chinese Academy of Sciences (Y60001RA01, Y80002RA01)

**Conflict of Interest:** The authors declared no potential conflicts of interest with respect to the research, authorship, and/or publication of this article.

**Ethical approval:** All procedures performed in this study were in accordance with the ethical standards of the institutional research committee and with the 1964 Helsinki Declaration and its later amendments. For this type of study, formal consent is not required.

**References**

1. Swanson, E.A., and Fujimoto, J.G. (2017). The ecosystem that powered the translation of OCT from fundamental research to clinical and commercial impact [Invited]. Biomed. Opt. Express 8, 1638–1664.

2. Varma, R., Bressler, N.M., Doan, Q.V., Gleeson, M., Danese, M., Bower, J.K., Selvin, E., Dolan, C., Fine, J., Colman, S., and Turpcu, A. (2014). Prevalence of and risk factors for diabetic macular edema in the United States. JAMA Ophthalmol. 132, 1334–1340.

3. Wong, W.L., Su, X., Li, X., Cheung, C.M., Klein, R., Cheng, C.Y., and Wong, T.Y. (2014). Global prevalence of age-related macular degeneration and disease burden projection for 2020 and 2040: a systematic review and meta-analysis. Lancet Glob. Health 2, e106–e116.

4. Gulshan, V., Peng, L., Coram, M., Stumpe, M.C., Wu, D., Narayanaswamy, A., Venugopalan, S., Widner, K., Madams, T., Cuadros, J., et al. (2016). Develop- ment and Validation of a Deep Learning Algorithm for Detection of Diabetic Retinopathy in Retinal Fundus Photographs. JAMA 316, 2402-2410

5. Li Z , He Y , Keel S , et al. (2018) Efficacy of a Deep Learning System for Detecting Glaucomatous Optic Neuropathy Based on Color Fundus Photographs. Ophthalmology. 125(8):1199-1206.doi:10.1016/j.ophtha.2018.01.023

6. Brown JM , Campbell JP , Beers A , et al. (2018) Automated Diagnosis of Plus Disease in Retinopathy of Prematurity Using Deep Convolutional Neural Networks. JAMA Ophthalmol. 136(7): 803-810.doi: 10.1001/jamaophthalmol.2018.1934

7. Kermany D S, Goldbaum M, Cai W, et al. Identifying Medical Diagnoses and Treatable Diseases by Image-Based Deep Learning[J]. Cell, 2018, 172(5):1122–1131.e9.

8. Ting DSW, Liu Y, Burlina P, Xu X, Bressler NM, Wong TY. AI for medical imaging goes deep. NatMed. 2018;24(5):539-540. doi:10.1038/s41591-018- 0029-3 16.

9. McCallister, E., Grance, T., & Scarfone, K. A. (2010). SP 800-122. Guide to Protecting the Confidentiality of Personally Identifiable Information (PII).

10. Goodfellow, I.J., Pouget-Abadie, J., et. al.: Generative adversarial networks. NIPS (2014).

11. Assessment of Deep Generative Models for High-Resolution Synthetic Retinal Image Generation of Age-Related Macular Degeneration. JAMA Ophthalmol. 2019 Mar 1;137(3):258-264. doi: 10.1001/jamaophthalmol.2018.6156.

12. Karras, T., Aila, T., et. al.o: Progressive growing of gans for improved quality, stability, and variation. arXiv (2017).

13. T. Zhang et al., SkrGAN: Sketching-rendering Unconditional Generative Adversarial Networks for Medical Image Synthesis, MICCAI 2019

14. Szegedy C , Vanhoucke V , Ioffe S , et al. (2016) Rethinking the Inception Architecture for Computer Vision. 2016 IEEE Conference on Computer Vision and Pattern Recognition (CVPR). 2818-2826.doi:10.1109/CVPR.2016.308

15. Simonyan K, Zisserman A. (2014) Very Deep Convolutional Networks for Large-Scale Image Recognition. CoRR. abs/1409.1556.https://arxiv.org/abs/1409.1556. Accessed November 11 2018.

16. Ting DSW, Cheung CY, Lim G, et al. Development and validation of a deep learning system for diabetic retinopathy and related eye diseases using retinal images from multiethnic populations with diabetes. JAMA. 2017;318(22):2211-2223. doi:10. 1001/jama.2017.18152 5.

17. Brown JM, Campbell JP, Beers A, et al; Imaging and Informatics in Retinopathy of Prematurity (i-ROP) Research Consortium. Automated diagnosis of plus disease in retinopathy of prematurity using deep convolutional neural networks. JAMA Ophthalmol. 2018;136(7):803-810. doi:10.1001/ jamaophthalmol.2018.1934 7.

18. Burlina P, Pacheco KD, Joshi N, Freund DE, Bressler NM. Comparing humans and deep learning performance for grading AMD: A study in using universal deep features and transfer learning for automated AMD analysis. Comput Biol Med.2017; 82:80-86. doi:10.1016/j.compbiomed.2017.01.018

19. Mahapatra D, Bhavna A, Suman S, Rahil G. Deformable medical image registration using generative adversarial networks. In: 2018 IEEE 15th International Symposium on Biomedical Imaging (ISBI 2018). Piscataway, NJ: IEEE; 2018: 1449-1453.

20. Baur C, Albarqouni S, Navab N. MelanoGANs: High Resolution Skin Lesion Synthesis with GANs. Preprint. Published online Month Day, 2018. arXiv. 1804.04338. 30.

21. Nie D, Trullo R, Lian J, et al. Medical Image Synthesis with Deep Convolutional Adversarial Networks. IEEE Transactions on Biomedical Engineering, 2018: https://ieeexplore.ieee.org/ stamp/stamp.jsp?arnumber=8310638. Accessed November 19, 2018

22. Daniel S.W. Ting, Lily Peng, Avinash V. Varadarajan, Pearse A. Keane, Philippe M. Burlina, Michael F. Chiang, Leopold Schmetterer, Louis R. Pasquale, Neil M. Bressler, Dale R. Webster, Michael Abramoff, Tien Y. Wong, Deep learning in ophthalmology: The technical and clinical considerations, Progress in Retinal and Eye Research, Volume 72, 2019, 100759, ISSN 1350-9462, https://doi.org/10.1016/j.preteyeres.2019.04.003.

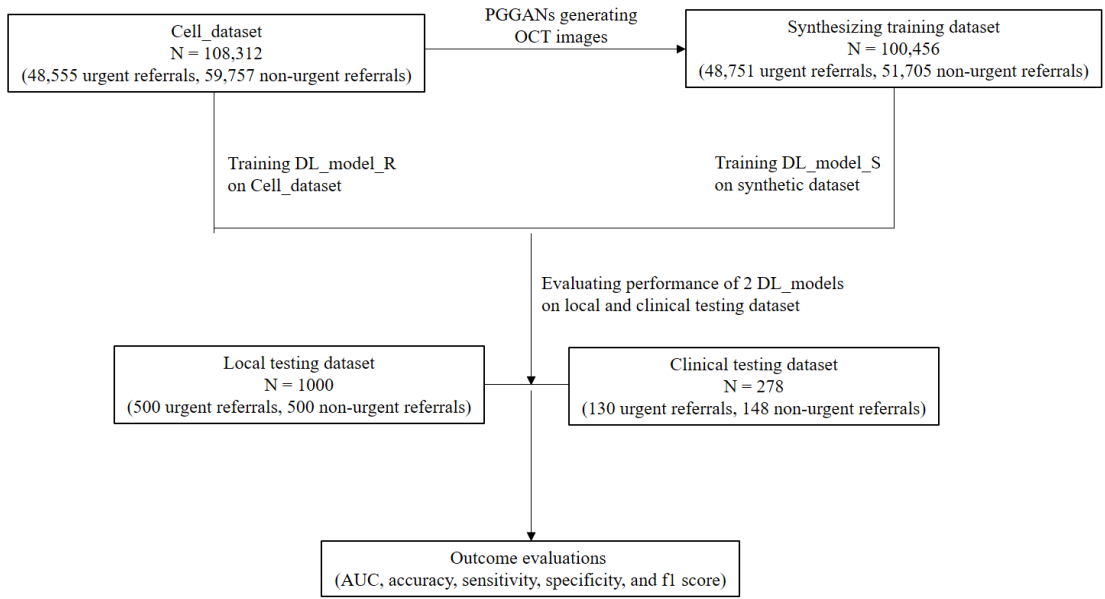

Figure 1 Assessment workflow of using the synthetic OCT images in the classification of various retinal disorders.

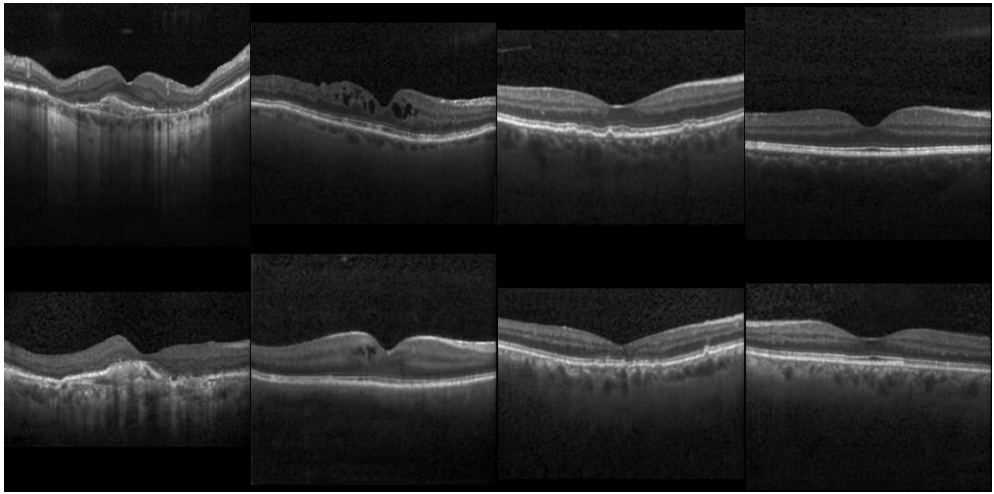

Figure 2 Examples of the synthetic OCT images (The real OCT images were above and the synthetic OCT images was below).

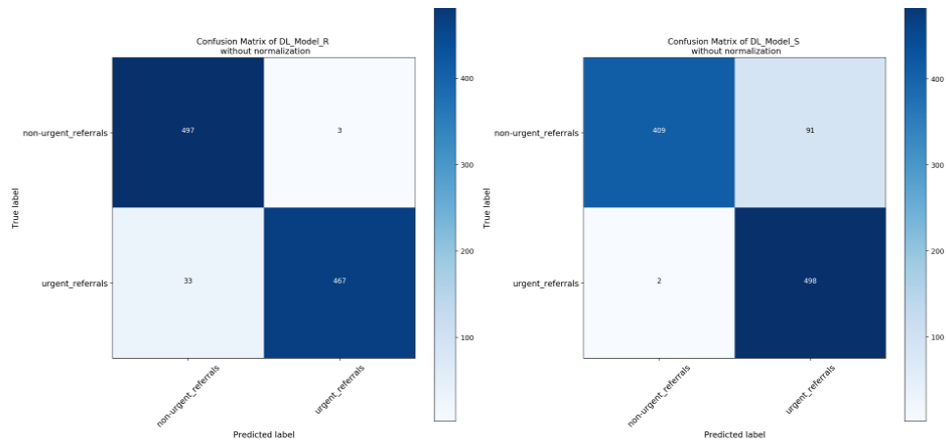

Figure 3 Confusion matrix of 2 DL models testing in local Cell validation dataset.

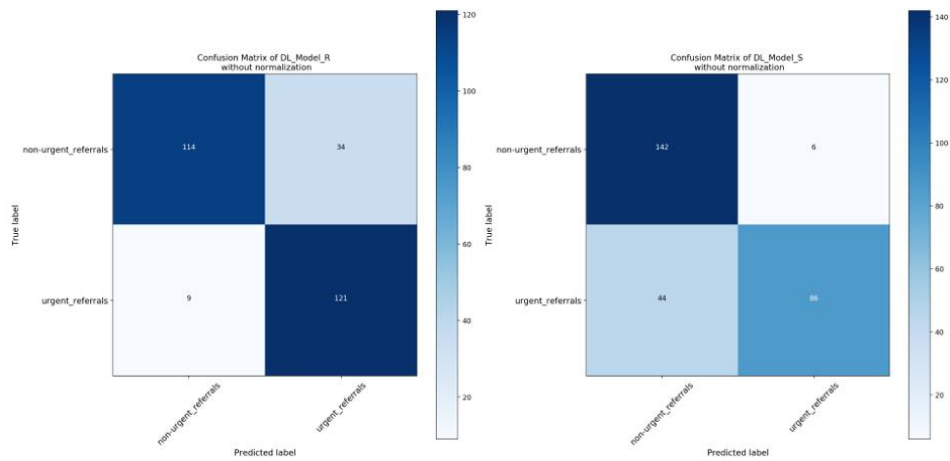

Figure 4 Confusion matrix of 2 DL models testing in clinical validation dataset.

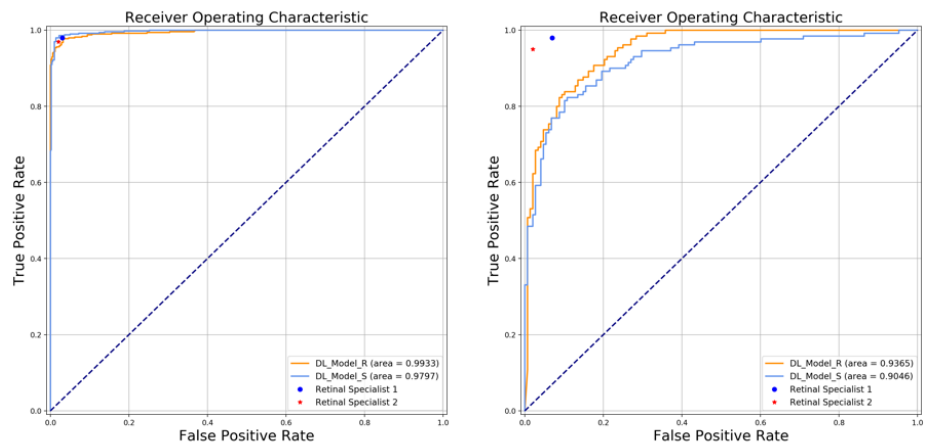

Figure 5 ROC curves of the two DL models tested in the local Cell validation dataset (left) and clinical validation data set (right).

Table 1. The Proportion of Poor Quality for Real and Synthetic OCT Images

|  | Poor Image Quality (%) |
|---|---|
| Retina Specialist 1 |  |
|   All | 7 (1.75%) |
|   Real | 3 (1.5%) |
|   Synthetic | 4 (2%) |
| Retina Specialist 2 |  |
|   All | 9 (2.25%) |
|   Real | 4 (2%) |
|   Synthetic | 5 (2.5%) |

Table 2. The Diagnostic Performance of 2 DL_Models and Retinal Specialists Testing in Local and Clinical Validation Datasets

|  | AUC (95% CI) | Precision (95% CI) | Recall (95% CI) | f1-score (95% CI) |
|---|---|---|---|---|
| A: Testing in Local Cell_Validation Dataset | | | | |
| Deep Learning Models | | | | |
| DL_Model_R | 0.99 (0.99 to 1.00) | 0.96 (0.95 to 0.98) | 0.96 (0.95 to 0.98) | 0.96 (0.95 to 0.97) |
| DL_Model_S | 0.98 (0.97 to 0.99) | 0.92 (0.90 to 0.94) | 0.91 (0.89 to 0.93) | 0.91 (0.89 to 0.93) |
| Human Experts | | | | |
| Retina Specialist 1 | 0.870 (0.839 to 0.901) | 0.915 (0.889 to 0.941) | 0.817 (0.781 to 0.853) | 0.863 (0.831 to 0.895) |
| Retina Specialist 2 | 0.827 (0.791 to 0.862) | 0.852 (0.819 to 0.885) | 0.794 (0.756 to 0.831) | 0.822 (0.786 to 0.857) |
| B: Testing in Clinical Dataset | | | | |
| Deep Learning Models | | | | |
| DL_Model_R | 0.94 (0.91 to 0.97) | 0.86 (0.82 to 0.90) | 0.85 (0.81 to 0.89) | 0.84 (0.80 to 0.88) |
| DL_Model_S | 0.90 (0.87 to 0.94) | 0.84 (0.80 to 0.88) | 0.82 (0.78 to 0.87) | 0.81 (0.76 to 0.86) |
| Human Experts | | | | |
| Retina Specialist 1 | 0.870 (0.839 to 0.901) | 0.915 (0.889 to 0.941) | 0.817 (0.781 to 0.853) | 0.863 (0.831 to 0.895) |
| Retina Specialist 2 | 0.865 (0.833 to 0.897) | 0.861 (0.829 to 0.894) | 0.870 (0.839 to 0.901) | 0.866 (0.834 to 0.897) |